# Modeling size controlled nanoparticle precipitation with the co-solvency method by spinodal decomposition

Simon Keßler[1], Friederike Schmid[2], Klaus Drese[1]

1: Fraunhofer ICT-IMM, Carl-Zeiss-Str. 18-20 55129 Mainz
2: Institut für Physik, Johannes Gutenberg-Universität Mainz, D55099 Mainz

**Abstract:** The co-solvency method is a method for the size controlled preparation of nanoparticles like polymersomes, where a poor co-solvent is mixed into a homogeneous copolymer solution to trigger precipitation of the polymer. The size of the resulting particles is determined by the rate of co-solvent addition. We use the Cahn-Hilliard equation with a Flory-Huggins free energy model to describe the precipitation of a polymer under changing solvent quality by applying a time dependent Flory-Huggins interaction parameter. The analysis focuses on the characteristic size $R$ of polymer aggregates that form during the initial spinodal decomposition stage, and especially on how $R$ depends on the rate $s$ of solvent quality change. Both numerical results and a perturbation analysis predict a power law dependence $R \sim s^{-\frac{1}{6}}$, which is in agreement with power laws for the final particle sizes that have been reported from experiments and molecular dynamics simulations. Hence, our model results suggest that the nanoparticle size in size-controlled precipitation is essentially determined during the spinodal decomposition stage.

## 1. INTRODUCTION

Because of their great potential in nano- and biotechnology, polymeric nanoparticles such as polymersomes have attracted growing interest during the last decades. [1,2] One possible application can be found in the field of drug delivery, where they serve as transport vehicles for medication. [3] A crucial property of such transport vehicles is their size, as it does not only determine their loading capacity, but also the composition of their protein corona in blood, which affects the retention times in the circulatory system. [4] Furthermore, the nanoparticle size plays a critical role in passive targeting of tumors, which is based on the Enhanced Permeability and Retention effect. [5]

A method to prepare polymersomes of a particular size is the co-solvency method or flash nanoprecipitation: A poor co-solvent is mixed into an initially homogeneous solution of a good solvent and a block copolymer to induce particle formation via self-assembly of the polymer. [6, 7] The co-solvency method can be implemented in different ways. A straightforward approach is to add co-solvent by drop injection to a polymer solution. In this method the size of the produced nanoparticles depends on the rate of co-solvent addition. [8] Thiermann et al. fed the co-solvent and the copolymer solution into continuous flow multilamination micro mixers and observed that the size of the synthesized particles decreases with an increasing flow rate, which connects nanoparticle size to an easily adjustable parameter of the experimental setup. [9] The micro mixer approach has several advantages as it yields narrower size distributions and can be done without additional steps like membrane extrusion to achieve acceptably low polydispersities.

A superficial comparison between the two realizations of the co-solvency method suggests that the size of the produced nanoparticles depends on a completely different quantity in both cases: the rate of co-solvent addition on the one hand and the flow rate on the other. However, due to the special design of multilamination micro mixers, an increase in flow rate decreases the mixing time of liquids that are fed into its inputs. [10] Changing the flow rate indirectly changes the mixing rate. Hence, in both cases particle sizes are found to depend on the mixing rates of the co-solvent and the copolymer solution.

To gain a deeper understanding of the size controlled preparation of polymeric nanoparticles with the co-solvency method, one must analyze how different rates of co-solvent addition affect the particle formation and how the particle size depends on the rate of co-solvent addition. In this article we consider the earliest stage of particle formation, the spinodal decomposition of an oversaturated polymer solution. We present a simple phase field model that can be used to determine the size of aggregates in that stage. The model is an extension of the popular Cahn-Hilliard model [11] for the dynamics of phase separation, where we represent the effect of mixing solvent and co-solvent in an effective manner by using time dependent interaction parameters. Using numerical simulations of an idealized mixing process, we show that the model can reproduce the dependence of particle size on mixing speed observed experimentally.

Our article is organized as follows: We present the model in section 2 and describe the simulation method in in section 3. In section 4, we explain the evaluation method, show our simulation results, compare them to a perturbation theory, and discuss scaling laws both have in common. These scaling laws are then compared to experimentally observed power laws in section 5. We summarize and conclude in section 6.

**2. THEORETICAL MODEL**

Experimentally, three different components are involved in the co-solvency method of nanoparticle synthesis: A polymer, a good solvent and a poor or selective co-solvent. The precipitation of the polymer is triggered and influenced by the continuous addition of co-solvent into the polymer solution – i.e., by solvent mixing. We assume that solvent mixing is fast on the time scales of the polymer phase separation and that the main effect of solvent mixing is a change of 'mean solvent quality' from 'good' to 'poor'. Thus, we incorporate solvent mixing by only taking into account the change in solvent quality: The three component system from the experiment is modeled by a two component system containing a polymer and only one solvent, which changes its quality over time. More specifically, at any given time we describe the momentary solvent mixture by one single effective solvent with an associated interaction parameter $\chi$ at a polymer-solvent contact. The addition of co-solvent into the solvent mixture is modeled by a temporal increase of $\chi$.

Close to a homogeneous ground state isothermal phase separation of incompressible binary mixtures in a fixed volume can be modeled by the Cahn-Hilliard equation. It describes the local evolution of a globally conserved dimensionless composition field $u$, for example the volume fraction of one component. The Cahn-Hilliard Equation is a special case of the generalized diffusion equation

$$\frac{\partial u}{\partial t} = -\nabla \cdot \left(-v\zeta M(u) \cdot \nabla \frac{\delta F}{\delta u}\right). \qquad (1)$$

$\zeta$ represents the scale of the mobility and $M(u)$ describes its dependence on the composition. $v$ is the volume of a polymer or solvent segment. Here we will use the "degenerate mobility" $M(u) = u \cdot (1 - u)$, which is suitable for the description of composition currents in incompressible mixtures. [12, 13] $\delta F/\delta u$ is the functional derivative of the free energy functional and can be interpreted as a chemical potential. With the free energy functional $F$ proposed by Cahn and Hilliard [11],

$$F = \frac{1}{v}\int d^d r \left\{ f(u) + \tfrac{1}{2}\lambda(\nabla u)^2 \right\}, \qquad (2)$$

($d$ is the spatial dimension) one obtains

$$\frac{\delta F}{\delta u} = \frac{1}{v}\left(\frac{\partial f}{\partial u} - \lambda \Delta u\right). \qquad (3)$$

Here, $f(u)$ is the free energy per segment in a homogeneous system, $\tfrac{1}{2}\lambda(\nabla u)^2$ represents surface contributions, and $\lambda$ is the gradient energy parameter. Insertion of eq. (3) into eq. (1) yields the Cahn-Hilliard Equation

$$\frac{\partial u}{\partial t} = \nabla \cdot \left[ \zeta M(u) \cdot \nabla \left( \frac{\partial f}{\partial u}(u) - \lambda \Delta u \right) \right]. \quad (4)$$

In this article we specifically consider polymeric systems. Hence we choose $\zeta = D/k_B T$ and $\lambda = R_g^2 \cdot k_B T$ with the segment diffusion coefficient $D = D_p N$, the Boltzmann factor $1/k_B T$ and the radius of gyration $R_g$. $D_p$ is the diffusion coefficient of a polymer chain composed of $N$ segments. The expression for $\lambda$ is an approximation to the gradient energy parameter for a binary homopolymer solvent mixture, which holds in the weak segregation limit where concentration gradients are weak. [14] We also use

$$f(u) = k_B T \left( \frac{u}{N} \ln u + (1-u) \ln(1-u) + \chi \cdot u(1-u) \right), \quad (5)$$

which is free energy per segment from the Flory-Huggins solution theory, [15, 16] where $\chi$ is the Flory-Huggins Interaction Parameter mentioned earlier.

To describe phase separation during solvent mixing, we assume that $\chi$ in eq. (5) explicitly depends on time, i.e. $\chi = \chi(t)$, and increases from $\chi(0) = \chi_0$ to $\chi_{max}$. We choose $\chi_0$ to be the spinodal interaction parameter, which is the value of $\chi$ where a homogeneous system becomes unstable. It depends on the mean polymer volume fraction in the system,

$$\bar{u} = \frac{1}{V} \int_V u \, dV,$$

as well as $N$ and is defined by the condition $\frac{\partial^2 f}{\partial u^2}(\bar{u}, t = 0) = 0$, which yields

$$\chi_0 = \frac{1}{2} \left( \frac{1}{N\bar{u}} + \frac{1}{1-\bar{u}} \right). \quad (6)$$

$\chi_{max} > \chi_0$ is a constant and we consider a situation where $\chi(t)$ grows linearly as

$$\chi(t) = \begin{cases} \chi_0 + s \cdot t & t < t_{max} \\ \chi_{max} & t \geq t_{max} \end{cases} \quad (7)$$

with $t_{max} = (\chi_{max} - \chi_0)/s$.

In the following, all lengths will be given in units of $l_0 = \sqrt{\lambda/(k_B T)} = R_g$, all times in units of $t_0 = l_0^2/(\zeta k_B T) = R_g^2/D$, and all energies in units of $k_B T (R_g^d/v)$. Eq. (4) can then be rewritten as

$$\frac{\partial u}{\partial t} = \nabla \cdot \left[ M(u) \cdot \nabla \left( \frac{\partial f}{\partial u}(u, t) - \Delta u \right) \right], \quad (8)$$

and the derivative of the free energy becomes

$$\frac{\partial f}{\partial u}(u, t) = \frac{1}{N} \ln u - \ln(1-u) + \frac{1}{N} - 1 + \chi(t) \cdot (1 - 2u). \quad (9)$$

In experiments, polymersomes are typically formed from amphiphilic diblock copolymers and stabilized by the hydrophilic block. [8,9] The model presented in this article neither incorporates the stabilization effects from copolymers nor is it able to describe an internal structure of polymer aggregates, because eq. (5) is restricted to homopolymers. Thus, the equilibrium state will always correspond to macroscopic phase separation. However, simulations of more detailed models have shown that nanoparticle self-assembly is initially dominated by the formation of unstructured droplets, and that the number of droplets after the initial stage largely determines the final number and size of particles. [17] The system defined by eq. (5) represents the simplest model system that reproduces this early stage of particle assembly.

In this article we restrict our investigations to the very early stages of phase separation, where the first patterns in the concentration profile form and where the gradients in the composition field are still small, which motivates the application of equation (4) with our choice of $\lambda$. In the context of different possible mechanisms that lead to the formation of structured copolymer-nanoparticles, [17,18] we focus on the spinodal decomposition before the first micelles appear. In these very early stages, the self-assembly should be driven mainly by the energetically unfavorable interaction between the co-solvent and the co-solvent-phobic block of the polymer, which leads to typical 'Cahn-Hilliard-type' spinodal decomposition patterns in the concentration profiles. [19] The solvent-philic block of the polymer, which is often incompatible with the other one, is mainly responsible for internal structure formation in aggregates once they have formed. So if the internal structuring does not significantly change the size of an aggregate, the substitution of the copolymer by a homopolymer of its hydrophobic block might still yield approximate results for its size. We shall see below that eq. (5) is indeed sufficient to describe the relation between particle sizes and mixing rates in the early stages of mixing.

A very recent publication also shows that it is possible to produce homopolymer particles stabilized by surface charges. [20] Besides an experimental part it also contains Molecular Dynamics simulations, where solvent mixing is modeled by a time dependent strength of the repulsive force. They observe very similar scaling laws as we do. Another work from the same group applied time dependent repulsive forces to Dissipative Particle Dynamics simulations for copolymers. [21] Although they only slightly touched the issue of particle size dependence on mixing time their curves look also similar to ours and the ones from. [20] Thus, the model presented in this article reproduces important features observed in much more complex particle models. Its simple structure allows a perturbation treatment and we will see that typical scaling laws observed in our simulations, the Molecular Dynamics simulations and the experiments are already inherent in the perturbation theory, which might pave the way for semi analytical approaches. In addition, the phase field model allows to study slow mixing processes with characteristic mixing times in the range of milliseconds or more, whereas Molecular Dynamics simulations are limited to microseconds. [20]

There also exists a pinning effect of structures, [22] which is caused by viscoelasticity in systems with asymmetric molecular dynamics, i.e. polymer solutions. It might also affect particle sizes and there are models that incorporate this effect. [23] However, the present model does not, because pinning did not occur in the experiments [9] we aim to describe.

Since we focus on situations where the particle formation is a thermodynamically driven process (initiated by spinodal decomposition), which does not involve a thermally activated crossing of free energy barriers as in nucleation theory, we do not include thermal noise in our theoretical model, Eq. (1). This corresponds to the limit $v \to 0$ in Eqs. (1),(2) (thermal noise would scale with $\sqrt{v/R_g^d}$ ) and is also motivated by the fact that the relative thermal fluctuations are generally small in polymeric systems.

For simplicity, we call the first aggregates of well-defined shape that emerge during spinodal decomposition 'particles'. How we exactly define these particles and how we determine their size is described in the discussion in section IV.

## 3. SIMULATION METHOD

Eqs. (7) - (9) with $M(u) = u(1-u)$ constitute the model equations. The applied scheme is very similar to the one used by Zhu et al., [24] which is a first order time accurate pseudo spectral method, and any Fourier transform was calculated by the FFTW library. [25] The domains are boxes $[0, L_b)^d \times \mathbb{R}_0^+$ with $d = 2,3$ and periodic boundary conditions. As initial conditions we use uniformly distributed random perturbations in the interval $[\bar{u} - 0.001, \bar{u} + 0.001]$ generated by the mersenne twister. [26] All numerical results are averages over 5 simulation runs with different initial perturbations and we performed simulations in both 2 and 3 dimensions to check the influence of dimensionality. As it will turn out, the particular dimension plays only a minor role, which allows investigations of main dependencies in 2D to speed up computation times.

The adjustable physical parameters in the model are $N, s, \chi_{max}$ and $\bar{u}$. The slope s in eq. (7) parameterizes the rate of solvent quality change. It is varied to investigate the effect of different solvent mixing rates on phase separation, while the three remaining parameters are kept constant. In this article, we will mostly study a model with parameters set to $\bar{u} = 0.1$, $N = 14$, and $\chi_{max} = 2$. Simulations for more realistic parameters are shown in section V. In 2D we used $400 \times 400$ grid points with a lattice constant 0.25, and set the time step to 0.005.

Using that lattice constant assures that the spatial resolution does not limit the smallest particles we encounter in our simulations (which is the particle size at $\chi_{max}$). For constant interaction this problem could also be approached by a rescaling of the spatial coordinate that involves the quench depth. [27] However, this might introduce numerical artifacts that lead to unphysical pinning close to the spinodal. Even though these artifacts can be avoided by proper normalization, [28, 29] we did not scale the spatial coordinate by $\Delta\chi$ because in our case it depends on time, meaning the system size would exhibit a temporal change if we kept our lattice size and number of grid points constant as is customary in simulations. We should also note that we did not encounter any pinning artifact in our simulations either. In 3D we used 64x64x64 grid points and a lattice constant of 1.

## 4. RESULTS AND DISCUSSION

4.1. Qualitative characterization of the demixing process

In all simulation runs the phase separation proceeds in a similar manner than in the case of constant interaction parameters $\chi$. In the first stage, termed spinodal decomposition, a bicontinuous pattern emerges in the concentration profiles and initially grows and coarsens on a relatively fast time scale, until droplets with well-defined interfaces have formed (see examples in Fig. 2). In the second stage, called Ostwald ripening, the droplet pattern coarsens on a very slow time scale. As stated at the end of section 2, we focus on the spinodal decomposition stage. This restriction requires us to identify the crossover time between spinodal decomposition and Ostwald ripening. To this end, we use a procedure proposed by Sofonea and Mecke, which is based on Minkowski measures.[30] Minkowski measures are a complete set of additive motion-invariant measures for unions of convex sets. Each measure assigns one real number to any polymer volume fraction profile depending on its morphology. Since the morphology of concentration profiles during phase separation depends on time, the Minkowski measures also do. One of these measures, from here on denoted by $C$, is the total boundary length of the union over all subsets in $[0, L_b)^d$ where the polymer volume fraction $u$ exceeds a predefined threshold $u_{th}$. During spinodal decomposition, polymer aggregates form on a fast time scale leading to a rapid temporal increase of $C$ and during Ostwald ripening, polymer aggregates merge on a large time scale leading to a slow decrease of $C$. These two characteristic regimes can be seen in Fig. 1, which shows $C$ as a function of time for two examples discussed below. The regimes are separated by a maximum of $C$ and the corresponding time is called the transition time $t_{tr}$.[30] Therefore, spinodal decomposition dominates for $t < t_{tr}$ and Ostwald ripening for $t > t_{tr}$. The remaining Minkowski measures yield equivalent estimates for the transition time.[30] To calculate the Minkowski measures we use the algorithm proposed by Mantz et al.[31]

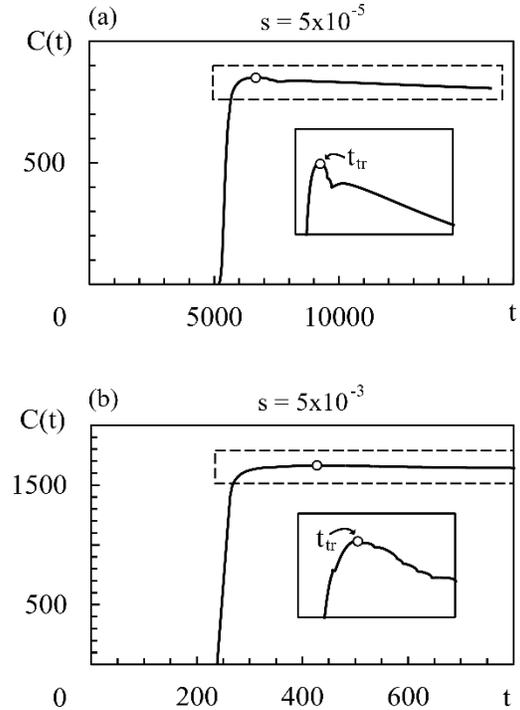

**FIG. 1**: Time series of the Minkowski measure $C$ (see text) for slopes $s = 5 \times 10^{-3}$ (a) and $s = 5 \times 10^{-5}$ (b). The time when $C$ reaches its maximum is defined as the transition time. The threshold value $u_{th}$ is set to 0.3 .

In the following, we first discuss exemplarily the effect of a time dependent interaction parameter on spinodal decomposition by comparing the results from $s = 5 \times 10^{-5}$ and $s = 5 \times 10^{-3}$. Fig. 1 shows the corresponding time series of $C$. The transition time obviously depends on $s$. Hence, spinodal decomposition happens faster for large values of $s$. Fig. 2 illustrates how different values of $s$ affect the morphology of the polymer volume fraction profiles during spinodal decomposition. The upper panel (Fig. 2 (a), (c), (e)) and the lower panel (Fig. 2 (b), (d), (f)) show temporal evolutions of the same initial polymer volume fraction profile for different growth rates $s$. At $t = 10$ the volume fraction profiles look very similar (Fig. 2 (a) and (b)). At $t = 200$, however, they deviate significantly from each other (Fig. 2 (c) and (d)) and at the end of the spinodal decomposition there are significantly smaller droplets for the larger quench rate (Fig. 2 (e) and (f)). Hence, we see that the time dependence in the interaction parameter does not only affect transition times but also the length scales of structures during spinodal decomposition. We can rationalize this observation as follows: With increasing $\chi$, one reaches deeper into the miscibility gap and the characteristic wavelength of the most unstable mode decreases. If $\chi(t)$ increases very slowly, the initially unstable modes have time to grow and dominate also the later stages of demixing. If $\chi(t)$ increases more rapidly, modes with smaller wavelengths take over and determine the final structure. Indeed, Fig. 2 demonstrates that the characteristic length scale of patterns in the initial stage of demixing (Fig 2 (c) and (d)) is larger than the characteristic length scale of the final droplet pattern (Fig. 2 (e) and (f)). We will analyze this effect at a more quantitative level further below in Section 4.3.

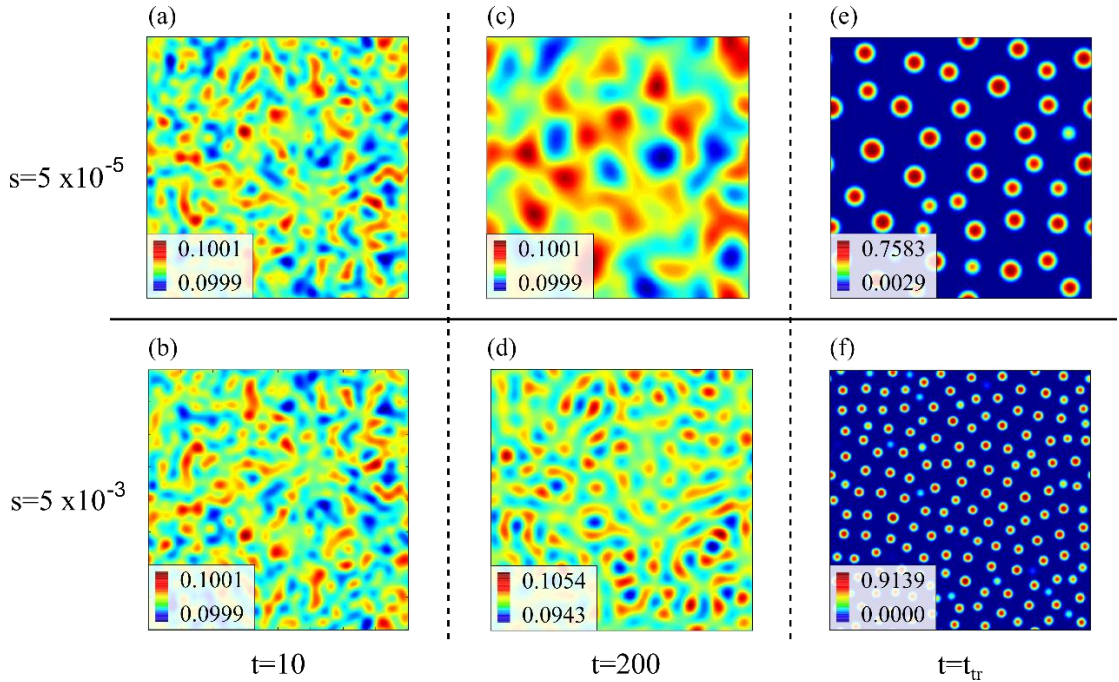

**FIG. 2:** Spatial distribution of polymer volume fraction $u(x, y, t)$ (color coded) in the domain $[0, L_b)^2$ at different times $t$ ($t=10$, $t=200$, and $t = t_{tr}$) during spinodal decomposition for $s = 5 \times 10^{-5}$ (upper panel: (a), (c), (e)) and $s = 5 \times 10^{-3}$ (lower panel: (b), (d), (f)). The transition time $t_{tr}$ depends on s and marks the time at which the behavior crosses over from spinodal decomposition to Ostwald ripening. The color coding is different for every snapshot and chosen such that the smallest value is blue and the largest dark red.

In general, bicontinuous patterns are favored for a larger range of composition variables $\bar{u}$ in 3D than in 2D, but for our set of parameters we observed droplets in both dimensions.

4.2. Definition of particles and quantitative determination of particle size

To examine the length scales in the volume fraction profiles more quantitatively, we evaluate the normalized radially averaged structure factor [24]

$$S(k,t) = \frac{S_c(k,t)}{n^d (<u^2> - <u>^2)}$$

where $S_c$ is the absolute value from the radial average of

$$\mathbf{S}(\vec{k},t) = \sum_{\vec{r},\vec{r}\prime} e^{-i\vec{k}\cdot\vec{r}\prime}[u(\vec{r}_{rsw}+\vec{r}\prime,t)\cdot u(\vec{r},t)- <u>^2].$$

Here $\mathbf{S}$ is the discrete Fourier transform of composition correlations in $d$ dimensions with wave vectors $\vec{k} \in \left\{\frac{2\pi}{L_b}\left(-\frac{n}{2}+1\right),\ldots,\frac{2\pi}{L_b}\frac{n}{2}\right\}^d$ and the summation is carried out over all grid points $\vec{r}, \vec{r}\prime$. $n$ is the number of grid points per site. The quantity $S_c(k,t)$ is calculated by averaging $\mathbf{S}(\vec{k},t)$ over the discs $\{\vec{k}: |\vec{k}| \in [k, k+2\pi/L_b]\}$, and $<\cdot>$ denotes the mean over the grid. The maximum and the first moment of the structure factor,

$$k_1(t) := \frac{\sum_k k\,S(k,t)}{\sum_k S(k,t)}, \quad (10)$$

are usually used to quantify a characteristic inverse length scale. We define the polymer aggregates at transition time as 'particles' and estimate their radius with

$$l_1 := \gamma\, 2\pi/k_1(t_{tr}) \quad (11)$$

and

$$l_{max} := \gamma\, 2\pi/k_{max}(t_{tr}) \quad (12)$$

where $k_{max}(t_{tr}) := \arg\max_k S(k,t_{tr})$ and $\gamma := 1/4$. We use two estimators since both $k_1$ and $k_{max}$ are reasonable choices to quantify a characteristic inverse length scale and we want to assess the difference between the two. The particle radius is thus taken to be one fourth of the characteristic wave length $2\pi/k_1(t_{tr})$ or $2\pi/k_{max}(t_{tr})$, respectively.

In addition, we used a standard image labeling algorithm to determine the total particle number $n_p$ and for each particle we calculate its sphere equivalent radius by

$$R_i := \begin{cases} \left(\frac{3V}{4\pi}\right)^{\frac{1}{3}}, & 3D \\ \sqrt{\frac{A}{\pi}}, & 2D \end{cases}$$

with $V$ and $A$ being the volume or the area of particle $i$. As a measure for the mean particle size we use the mean radius

$$R := \frac{1}{n_p}\sum_i R_i. \quad (13)$$

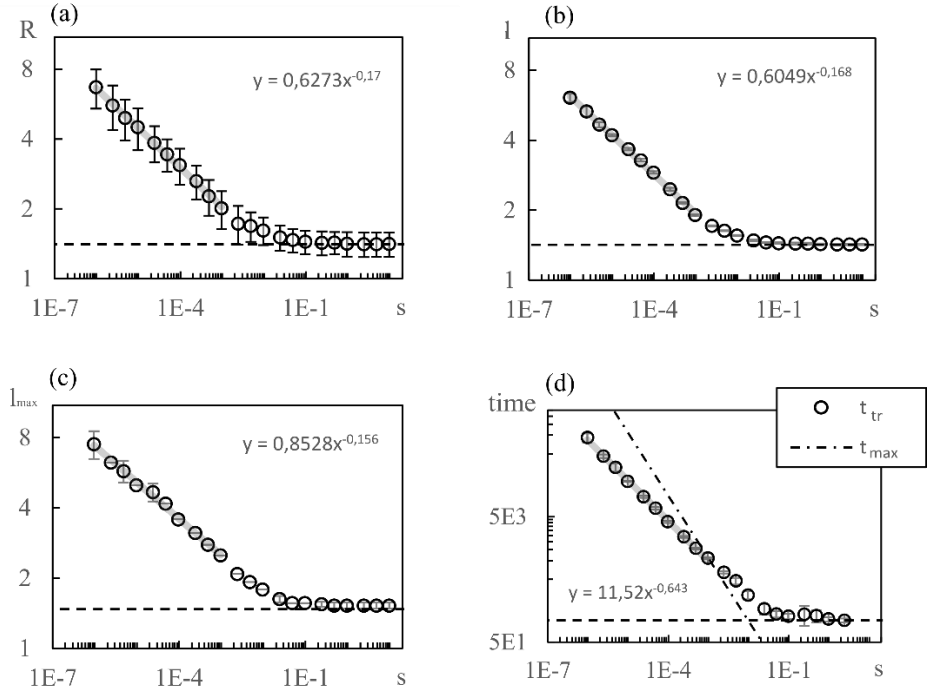
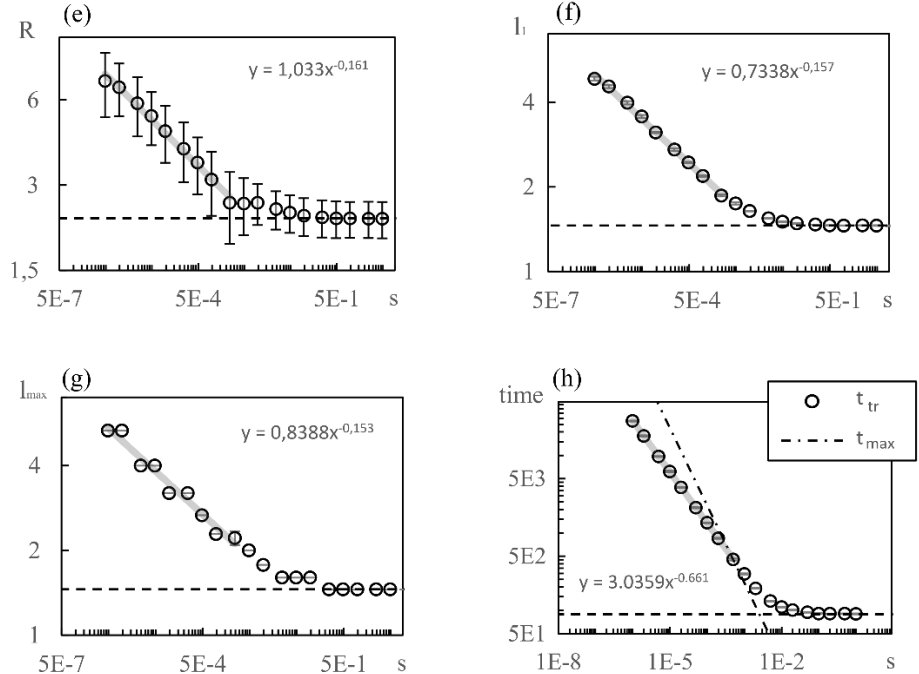

**FIG. 3:** Particle size ((a), (b), (c)) and transition time with $t_{max}$ (d) vs. solvent mixing rate $s$ in double logarithmic representation for 2D. Error bars in (b)-(d) indicate the standard deviation over 5 simulation runs. In (a) the error bars indicate the mean standard deviation of droplet radii within a specific simulation run to indicate polydispersity. The statistical variance of $R$ is similar to $l_1$ and $l_{max}$ and is not shown. The dashed horizontal lines give simulation results for $\chi(t) = \chi_{max}$ and the grey solid lines are regression lines (discussed in IV.C.3). (e)-(f) show the same as (a)-(d) but for 3 dimensions.

It should be emphasized that we have to restrict to the very early stages of particle formation, where no sharp interfaces are present, if we use the Cahn-Hilliard Equation. So actually we are interested in structures as they appear for example at $t = 200$ in fig. 2 (d) but since it is hard to define a clear measurement specification in the early stages we pick the particles at transition time as representatives because the structures from earlier times seem to imprint onto them.

4.3. Dependence of particle size and transition time on solvent mixing rate

Fig. 3 shows the simulation results for $R$, $l_1$, $l_{max}$ and the transition time $t_{tr}$ as a function of the mixing rate $s$. $l_1$, $l_{max}$, and $R$ take slightly different values but the progression of their data points is the same. All simulation results decrease monotonically in $s$ and show the same asymptotic behavior for large $s$.

We begin with discussing the asymptotic behavior. Since eq. (7) indicates that the value of the time dependent interaction parameter in the limit of infinitely fast solvent mixing is given by

$$\lim_{s \to \infty} \chi(t) = \begin{cases} \chi_0, & t < 0 \\ \chi_{max}, & t \geq 0 \end{cases},$$

the asymptotes are expected to correspond to the simulation results for an instantaneous quench with constant interaction parameter $\chi_{max}$. To check this assumption the corresponding results are represented by the dashed horizontal lines in Fig. 3. The data points clearly converge to these lines, hence the asymptotes are consistent with the expectations. In our set of equations, (7) - (9), a constant interaction parameter is achieved by substituting eq. (7) by $\chi(t) = \chi_{max}$. Next we define an asymptotic regime and discuss which values of $s$ belong to that regime. The growth of $\chi(t)$ is cut off when $t$ becomes greater than $t_{max}$. We call the asymptotic regime the values of $s$ for which the choice of the cutoff affects the simulation results at the transition time. This is clearly the case if $t_{tr}$ exceeds $t_{max}$. Hence the condition

$$t_{tr} > t_{max}(s) = \frac{\chi_{max} - \chi_0}{s}, \quad (14)$$

defines the asymptotic regime. The function $t_{max}(s)$ is represented by the dash-dotted line in Fig. 3 (d) and (h). It crosses the simulation results for $t_{tr}$ at $s \approx 2 \cdot 10^{-4} - 5 \cdot 10^{-4}$ in both 2D and 3D. The complement of the asymptotic regime is called non-asymptotic regime.

The numerical results for particle size in Fig. 3 show a remarkable similarity to the behavior of structure sizes that occur during continuous cooling of an alloy, which was investigated with a perturbation theory long time ago, including a typical scaling law with an exponent $-1/6$.[32] This similarity does not come as a surprise due to the formal relation of the underlying models. We are going to verify the scaling laws in a semi analytical manner and check if the actual values of the data points agree with this theory and not only their qualitative progression. To this end, we first expand eq. (8) in $u$ about the homogeneous ground state with a mean polymer volume fraction $\bar{u}$. After a Fourier transform in space, we obtain the ordinary differential equations

$$\frac{\partial c_{\vec{m}}}{\partial t}(t) = a(k, t) \cdot c_{\vec{m}}(t), \quad (15)$$

where $c_{\vec{m}}$ are the Fourier coefficients of the perturbation $(u(\vec{r}, t) - \bar{u})$, and $a(k, t)$ is given by

$$a(k, t) := -k^2 \cdot \left[ k^2 + \frac{\partial^2 f}{\partial u^2}(\bar{u}, t) \right]$$

with $k = 2\pi \sqrt{\sum_{j=1}^{d} m_j^2}/L_b$ and $m_j \in \mathbb{Z}$. The solution to eq. (15) reads $c_{\vec{m}}(t) = A_{\vec{m}} \cdot \exp(\int_0^t a(k, t') \, dt')$ and we define

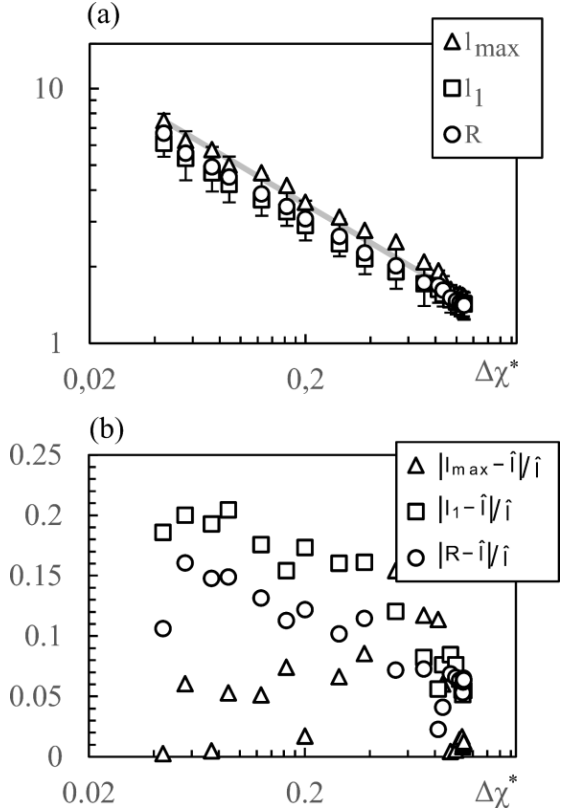

**FIG. 4:** (a): Characteristic particle size vs. parameter $\Delta\chi^*$. The symbols correspond to simulation data shown in Fig. 3 (a) to (c), the line represents the prediction of the linear approximation $l^*$. (b) Relative deviation between simulation data and $l^*$. The data for 3D is not shown but looks very similar.

$$k^* := \arg\max_k \left( \int_0^{t_{tr}} a(k, t') \, dt' \right).$$

Inserting eqs. (9) and (7) into $a(k, t)$ gives

$$k^* = \begin{cases} \sqrt{\dfrac{s\, t_{tr}}{2}} & t_{tr} < t_{max} \\ \sqrt{\left(1 - \dfrac{t_{max}}{2\, t_{tr}}\right)(\chi_{max} - \chi_0)} & t_{tr} \geq t_{max} \end{cases}. \quad (16)$$

We use $k^*$ to estimate the particle radius within the linearized theory. The corresponding estimator is

$$l^* := \gamma \frac{2\pi}{k^*}$$

which is defined analogously to eqs. (11) and (12). The comparison between the numerical results and $l^*$ is depicted in Fig. 4. The data points are identical to the data points in Fig. 3 (a) to (c) but they are plotted in a different representation, namely versus $\Delta\chi^* := k^{*2}$ instead of $s$. In this representation, $l^*$ becomes

$$l^* = \gamma \frac{2\pi}{\sqrt{\Delta\chi^*}}, \quad (17)$$

which is shown as a straight line in Fig. 4 (a). The data points in the asymptotic regime in Fig. 3 collapse onto a single accumulation point at $\Delta\chi^* = \lim_{s\to\infty} k^{*2} = \chi_{max} - \chi_0$ because $\lim_{s\to\infty} t_{max} = 0$ while $t_{tr}$ has a lower bound greater than zero (cf. Fig. 3 (d) and (h)). Fig. 4 shows that the prediction for the particle size from the linearized theory, $l^*$, approximates the numerical results with a relative deviation of less than 20 %. Regarding our interpretation of $l^*$, $l_{max}$ should be the best approximation and it can be seen from fig. 4 (b) that its deviation is even less than 10%. Hence, we conclude that the perturbation theory yields a good approximation to the numerical results at transition time. The scaling $l^* \propto \Delta\chi^{*-0.5}$ also reminds of the relation between particle size and quench depth for constant interaction parameters, [33,34] which gives $\Delta\chi^*$ the interpretation of an effective constant quench depth.

To establish a relation between $k^*$ and $s$ we plot $t_{tr}$ against $\Delta\chi^*$ in fig. 5 and observe that $t_{tr} \propto \Delta\chi^{*-2}$, which is also reminiscent of a scaling behavior for constant quench depths. The proportionality can be used to formulate approximate scaling laws for the non-asymptotic regime in Fig. 3. Inserting $t_{tr} \propto \Delta\chi^{*-2} = k^{*-4}$ into the case for $t_{tr} < t_{max}$ from eq. (16) yields

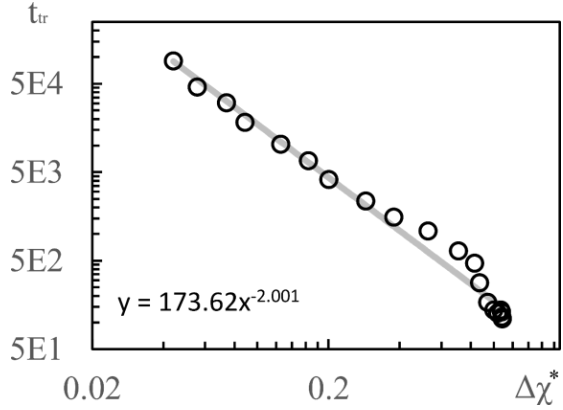

**FIG. 5:** (a): Transition times $t_{tr}$ from Fig. 3 (d) plotted vs $\Delta\chi^*$. The solid line is a regression line. The simulation results for 3D show a very similar scaling.

$$t_{tr} \propto s^{-\frac{2}{3}} \quad (18)$$

for $t_{tr} < t_{max}$. Employing the proportionality (18) into eq. (16) and combining it with definition of $l^*$ leads to

$$l^* \propto s^{-\frac{1}{6}} \quad (19)$$

for $t_{tr} < t_{max}$ or $s > \max\{s : t_{tr} < t_{max}\}$. The solid lines in Fig. 3 (a)-(c) and (e)-(g) are regression lines to the corresponding data points. Their equations are shown in the diagrams and their exponents deviate about 10 % and less from $-1/6$. Hence, the semi analytical approach verifies the predictions from the perturbation theory. The deviation of the slope in Fig. 3 (d) and (h) from $-2/3$ in eq. (18) is 3.5 % and less. Therefore, the scaling behavior of the numerical data in the non-asymptotic regime comes very close to the predicted scaling behavior from eqs. (18) and (19). These two equations are independent from $\bar{u}$ and $N$. The parameter $\chi_{max}$ affects $t_{max}$ and thus the extent of the non-asymptotic regime, but not the simulation results within that regime. Since the only independent parameters in the model other than $s$ are $\bar{u}$, $N$, $\chi_{max}$, the scaling laws in the non-asymptotic regime seem to be an universal feature – at least provided that different choices of $\bar{u}$ and $N$ do not destroy the analogy between linearly time dependent and constant interaction parameters. Even though they are not shown in the current publication we performed simulation runs for different parameters and the scaling laws were always observed with the same exponents within an error of 20% and less. Figure 4 implies that even the actual values of the particles sizes correspond to the perturbation theory.

## 5. REFERENCE TO EXPERIMENTS

From a practical point of view, the most relevant part in section 4 is the scaling law $l^* \propto s^{-\frac{1}{6}}$. Batch experiments with drop injection of selective solvent [8] report that the mean vesicle or micelle radius depends on the rate of co-solvent addition according to a power law with an exponent of approximately -0.13. The drop-wise co-solvent addition at a constant rate could imply the applicability of a linear time dependence of the interaction parameter allowing a direct comparison between -0.13 and -1/6, which is a good agreement. In experiments where nanoparticles are produced continuously inside micro mixers [9] it was also observed that the mean particle radius depends on the flow rate according to a power law with an exponent of -0.11, -0.13, or -0.17 depending on the mixer. For a comparison with the micro mixer approach, however, $s$ has to be translated into a flow rate $v$. Usually, the mixing time (corresponding to $t_{max}$ in our model) is inversely proportional to the Reynolds number and thus, to $v$. [10] So linear interpolations of the temporal co-solvent volume fraction evolutions in such a mixer show slopes proportional to $v$. This leads to scaling laws $l^* \propto v^{-\frac{1}{6}}$, which is also in good agreement with the experiments.

To make a more quantitative comparison we calculate mixing times $\tau$ for different flow rates in the Cater Pillar Micro Mixer with an analytical approach [35] and assume $s = (\chi_{max} - \chi_0)t_0/\tau$, with the time scale $t_0 = R_g^2/D$ defined in the beginning. This leads to

$$s \approx (\chi_{max} - \chi_0)\frac{R_g^2}{D}100\frac{min}{sec \cdot ml}v, \qquad (20)$$

where $R_g^2$ and $D$ have SI units and $v$ is is given in ml/min like in the experiments. The polymer $PB_{130}PEO_{66}$ possesses a molar mass of $M \approx 10 \, kg/mol$.[9] Unfortunately, the density for the copolymer was not measured but the homopolymer densities are $\rho_{PB} = 0.96 \, kg/l$ and $\rho_{PEO} = 1.2 \, kg/l$, so we estimated the copolymer density by their mean value, $\rho \approx 1.08 \, kg/l$. The polymer content in the dilute solution was about $c = 4 \, (g \, Polymer)/(l \, solvent)$. Basic algebra leads to a mean polymer volume fraction of $\bar{u} = \alpha/(1+\alpha)$ with $\alpha = c/\rho$. Using the values above we have $\bar{u} = 0.004$. Both the molar mass and the density of THF is comparable to the molar mass and the density of the monomers PB and PEO, resulting in similar molar volumes. Thus we estimated $N = 190$ and set $D$ to the diffusion coefficient of THF in water, which is about $10^{-9} m^2/s$. We substituted the PEO part by PB and estimated $R_g$ of the resulting homopolymer from its molar mass $M$ by a relation [36] which is valid for PB in THF and gives $R_g \approx 10 \, nm$.

Simulations were performed with $\bar{u} = 0.004$, $N = 190$ and $\chi_{max} = 16$. It should be noted that we Taylor expanded $\ln(u)$ in eq. (9) up to 10$^{th}$ order around $\bar{u}/10$ to avoid numerical difficulties caused by large $N$. Using $R_g$ and $D$ as mentioned above we converted $R$ to the nanometer scale and calculated flow rates with eq. (20). The results are shown in fig. 7. The open symbols are data from the experiments for symmetric flow conditions and the black dots represent our simulation results. CPMM, SIMM, SFIMM denote specific types of micro mixers and A and B refer to different end groups attached to the polymer. The SFIMM and SIMM [37,38,39] are pictured for the sake of completeness – strictly speaking $v$ is the corresponding flow rate in the CPMM. It can be seen that the model is able to reproduce both the scaling law and typical length and time scales of the experiments but it predicts roughly two times smaller particles. This could either be due to the rather rough approximations for $D$ and $R_g$, the application of an implicit sovent model [40], or the restriction to homopolymers. The final particle size is also influenced by 'technical' issues like the choice of $u_{th}$, so strict quantitative comparisons should be taken with care. It also should be emphasized that fig. 7 shows simulation results for homopolymers and experimental data for copolymers, i.e. components of very different composition. Comparing the experimental data for $PB_{130}PEO_{66} - H$ (sample A) and $PB_{130}PEO_{66} - CO - CH_2 - CH_2 - COOH$ (sample B) in the CPMM it can be seen that the composition of the polymer chain significantly shifts the data.

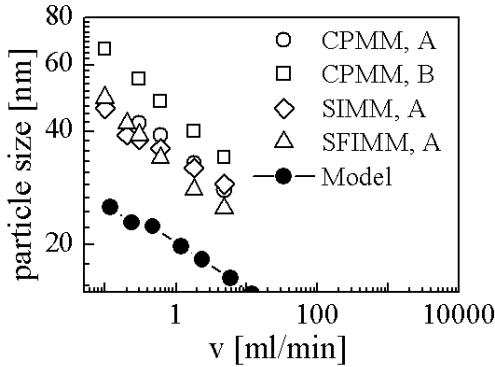

**FIG. 7:** Comparison between our model experiments from Thiermann et al. 9 CPMM, SIMM, SFIMM denote mixer types and A and B refer to different end groups of the polymer (see text for details) .

## 6. SUMMARY AND OUTLOOK

We have described nanoparticle precipitation by spinodal decomposition.

The simulations reproduce power laws as well as typical length scales for the size of vesicles and micelles from experiments. [8,9] These scaling laws are also in par with analytical investigations of spinodal decomposition during continuous cooling [32] and our results also agree with more complex particle models for homopolymer precipitation [20], where similar exponents were observed ($\approx -0.17$). Thus, the main result of the present article is that the thermodynamic notion of spinodal decomposition is a promising frame to study size controlled flash nanoprecipitation. Compared to particle models, field theories and especially phase field models require less computation time and grant access to time scales corresponding to mixing times in experiments. Computation time also benefits from the fact that the scaling laws can be investigated in 2D, since 2D and 3D simulations show the same behavior, which allows relatively efficient explorations of parameter spaces. Due to their simple structure even an analytical treatment in the frame of a perturbation theory might be possible.

Scaling laws $l \propto \alpha^{-\frac{1}{6}}$ were also found in a recent publication, which considered the structuring of polymer solutions in the spinodal area upon solvent evaporation [41], where $l$ is a typical structure size and $\alpha$ a constant evaporation rate. The authors added $\alpha$ as a source term in a Cahn-Hilliard-Cook equation. Within a typical "Flory-Huggins"-phase diagram with axes $\bar{u}$ and $\chi$, they advance into the spinodal area in the $\bar{u}$-direction, while we move in the $\chi$-direction. The fact that both processes yield the same scaling behavior suggests that the scaling should just depend on the distance to the spinodal line independent of the direction in the $\bar{u} - \chi$-plane.

As far as the comparison between homopolymers and copolymers in fig. 7 is concerned, a possible interpretation of the similar particle size behavior could be that the co-solvent addition controls the size of the vesicles mainly by determining the size of their micellar predecessors (cf. mechanisms I and II [17,18]) and that 'population balance effects' like flow induced collision-coagulation and break-up of aggregates in the micro channels might play a minor role, if any. Thus we have also identified one possible mechanism that determines the nanoparticle size in micromixers.

The similar behavior of homopolymer and copolymer particle size might also imply that the principal effect behind size controlled nanoparticle precipitation could be independent of the actual polymer architecture

In the future, we plan to couple solvent mixing to more sophisticated free energy models, [19,42] which are able to describe copolymers and the vesicle formation process, in order to capture the nanoparticle self-assembly also in the later stages of the aggregation process. Furthermore, it would be interesting to compare simulations for three component systems to our effective two component system and to analyze explicitly how the phase separation process depends on the time-dependent solvent composition. In our study, we have focused on liquid-liquid phase separation, where crystallization and solidification effects can be neglected. Recent experiments on semi-crystalline copolymers [43] have shown that the effect of solvent exchange (in this case, solvent evaporation) on the dynamics is very different if demixing interferes with solidification. For example, the characteristic length scales of the resulting structures no longer depend on the solvent evaporation rate, and the experiments can be described within a model based on homogeneous nucleation theory. In the future, it will also be interesting to consider the competition of liquid-liquid phase separation and solidification in more detail.

## ACKNOWLEDGEMENTS

The authors acknowledge funding from EFRE.

## REFERECNES